\documentclass[11pt]{article}
\usepackage{graphicx,amsmath,bm, amsthm,mathrsfs,amssymb, braket, verbatim}
\usepackage[usenames]{color}
\usepackage{ulem,mathtools}
\usepackage{authblk}
\usepackage{cite}
\usepackage{tcolorbox}
\usepackage[section]{placeins}

\newcommand{\ee}{\end{equation}}
\newcommand{\word}[1]{\,\,\mbox{#1}\,\,}
\newcommand{\reff}[1]{(\ref{#1})}
\newcommand{\beq}{\begin{equation}}
\newcommand{\eeq}[1]{\label{#1}\end{equation}}
\newcommand{\beqa}{\begin{eqnarray}}
\newcommand{\eea}{\end{eqnarray}}
\newcommand{\eeqa}[1]{\label{#1}\end{eqnarray}}
\newcommand{\beg}{\begin{equation*}}
\newcommand{\eeg}{\end{equation*}}

\newcommand{\bsplit}{\begin{split}}
\newcommand{\esplit}{\end{split}}

\usepackage{subfig} %To put many figures side by side
\usepackage{circuitikz} %To draw the electrical circuits
\usepackage[capposition=bottom]{floatrow} %To add notes under a figure
\usepackage{epigraph} %For quotes at the beginning of sections and text
\usepackage{appendix}
\usepackage{rotating}
\allowdisplaybreaks

\title{Non-singular vortices with positive mass in 2+1-dimensional Einstein gravity with AdS$_3$ and Minkowski background}

\author[]{Ariel Edery\thanks{aedery@ubishops.ca}}

\affil[]{Department of Physics and Astronomy, Bishop's University, 2600 College Street, Sherbrooke, Qu\'{e}bec, Canada, J1M 1Z7.\vspace{1em}}

\begin{document}
\date{}
\maketitle
\begin{abstract}
In previous work, black hole vortex solutions in Einstein gravity with AdS$_3$ background were found where the scalar matter profile had a singularity at the origin $r=0$. In this paper, we find numerically static vortex solutions where the scalar and gauge fields have a non-singular profile under Einstein gravity in an AdS$_3$ background. Vortices with different winding numbers $n$, VEV $v$ and cosmological constant $\Lambda$ are obtained. These vortices have positive mass and are not BTZ black holes as they have no event horizon. The mass is determined in two ways: by subtracting the numerical values of two separate asymptotic metrics and via an integral that is purely over the matter fields. The mass of the vortex increases as the cosmological constant becomes more negative and this coincides with the core of the vortex becoming smaller (compressed). We then consider the vortex with gravity in asymptotically flat spacetime for different values of the coupling $\alpha=1/(16 \pi G)$. At the origin, the spacetime has its highest curvature and there is no singularity. It transitions to an asymptotic conical spacetime with angular deficit that increases significantly as $\alpha$ decreases. For comparison, we also consider the vortex without gravity in flat spacetime. For this case, one cannot obtain the mass by the first method (subtracting two metrics) but remarkably, via a limiting procedure, one can obtain an integral mass formula. In the absence of gauge fields, there is a well-known logarithmic divergence in the energy of the vortex. With gravity, we present this divergence in a new light. We show that the metric acquires a logarithmic term which is the $2+1$ dimensional realization of the Newtonian gravitational potential when General Relativity is supplemented with a scalar field. This opens up novel possibilities which we discuss in the conclusion.  
\end{abstract}
%\thispagestyle{empty}
%\end{titlepage}
\setcounter{page}{1}
\newpage
\section{Introduction}\label{Intro}

Though vortices, non-perturbative topological solutions in $2+1$ dimensions, have been studied for a long time, the study of gravity's effect on them is somewhat more recent. This is partly related to the fact that Einstein gravity in $2+1$ dimensions is trivial in the sense that outside localized sources the vacuum spacetime is locally flat (though there are global effects \cite{Deser}). The BTZ black hole solution \cite{BTZ1,BTZ2} in an AdS$_3$ background revived considerable interest in $2+1$-dimensional Einstein gravity. Here was a black hole with a horizon (if you include rotation there is an inner and outer horizon), with thermodynamic properties but in contrast to black holes in $3+1$ dimensions had no curvature singularity. Later on, black holes with spherical scalar hair in Einstein gravity in an AdS$_3$ background were studied analytically \cite{Cadoni}. Besides a real scalar field, they also considered a complex scalar field with a potential, which allowed them to construct black hole vortex solutions. The scalar field profile for the vortex had a singularity at the origin $r=0$ and approached zero asymptotically. Though this zero asymptotic value is not the minimum of the potential it satisfies the Breitenlohner-Freedman (BF) bound \cite{BF} for the AdS$_3$ background (in fact it saturates the BF bound). In particular, they obtain a compact analytical expression for the black hole mass $M$ in terms of the scalar charge $c$ and the winding number $n$: $M=\lambda_2 \,c^2/4 +n^2$ where $\lambda_2$ is a parameter appearing in the potential. More recently, the effect of vortices on the tunneling decay of a symmetry-breaking false vacuum in $2+1$ dimensional Einstein gravity was studied \cite{Rich}. They found that the tunneling exponent for vortices, the dominant factor in the decay rate, is half that of Coleman-de Luccia bubbles \cite{Coleman}. In turn this implied that the vortices are short-lived which rendered them cosmologically significant. There have also been studies of black holes in $2+1$ dimensions in the theory of BHT massive gravity \cite{BHT}. Besides black holes in AdS space the authors find black holes in de Sitter space, regular gravitational solitons and kinks as well as wormhole solutions. 

As already mentioned, for the black hole vortex in an AdS$_3$ background discussed above, the scalar field was singular at the origin. In our work, we find numerically vortices with non-singular matter composed of gauge and complex scalar fields under Einstein gravity in an AdS$_3$ background. These are not black hole solutions as there is no event horizon. We obtain vortices of positive mass for different cosmological constant $\Lambda$, winding number $n$ and VEV $v$. We obtain an expression for the mass of the vortex in two different ways: via the subtraction of the asymptotic values of two metrics and via an integral over the scalar and gauge matter profiles only. The two must necessarily match providing a strong check on our numerical results. The integral mass formula can be expressed conveniently as $n^2 v^2$ multiplied by an integral over matter profiles that always plateau to unity (we therefore see a connection with the $n^2$ dependence of the mass obtained in \cite{Cadoni} mentioned above). The $n^2v^2$ factor is reflected in our numerical results. We observe that for $n=2$ and $v=2$ the masses are significantly larger than their $n=1$, $v=1$ counterparts. For these two cases, we also find that the metric near the core of the vortex departs significantly from its asymptotic ``$r^2$" dependence (in other words, near the origin, it looks nothing like the BTZ black hole metric). We also observe that the mass of the vortex increases as the cosmological constant becomes more negative and that this coincides with the vortex becoming more compressed (core smaller).  We obtain an analytical expression for the cosmological constant $\Lambda$ (assuming $\Lambda \ne 0$) in terms of the VEV $v$, the coupling constant $e$ between scalar and gauge fields, and a parameter $q$ that describes how the gauge field plateaus asymptotically to $n$. The expression for $\Lambda$ is negative so that de Sitter vortices, just like de Sitter black holes, do not appear to exist in $2+1$ dimensional Einstein gravity (in contrast, the BHT massive gravity \cite{BHT} mentioned previously supports de Sitter black holes.) For a given $\Lambda$, $e$ and $v$, we compare the analytical value of $q$ to the one obtained numerically and they match quite well. 

We then consider the vortex with gravity in an asymptotically Minkowski spacetime (zero cosmological constant) for different values of $\alpha=1/(16 \pi G)$. The mass formula, labeled $M_{flat}$, is in agreement with the deficit angle expected in an asymptotically flat $2+1$ dimensional spacetime \cite{Deser}. Asymptotically, we obtain a conical spacetime with a given angular deficit. As one approaches the origin and hence the core of the vortex, the Ricci scalar increases to its maximum value and there is no singularity. We therefore have a smooth spacetime that transitions from a region of highest curvature at the origin and near the core of the vortex to an asymptotically flat conical region with angular deficit. Both the curvature at the origin and the angular deficit increase significantly as $\alpha$ decreases. In contrast, the mass of the vortex hardly changes with $\alpha$. For comparison, we also consider the vortex without gravity in fixed Minkowski spacetime. For this case, the mass cannot be obtained by the method of subtracting two asymptotic metrics. However, remarkably, by taking the appropriate limit of $M_{flat}$, we obtain an integral mass formula for it that works perfectly well as it does not refer to metrics, Newton's constant or the cosmological constant. Though this integral looks completely different from the standard energy integral presented in quantum field theory texts, we show that they are indeed equivalent when the equations of motion are used. 

It is well known that in the absence of gauge fields, the standard vortex in fixed Minkowski spacetime has a logarithmic divergence in the energy \cite{Weinberg}. This persists in the presence of gravity. But with gravity, we can present this divergence in a new light. We show that the metric acquires asymptotically a term of the form $G \,m \,\ln(r)$. This logarithmic term can be viewed as the $2+1$ dimensional realization of the Newtonian gravitational potential with mass $m$ when General Relativity is supplemented with a complex scalar field and symmetry breaking potential. The parameter $m$ turns out to be proportional to $n^2v^2$, the same mass dependence previously mentioned above. We discuss the implications of this in the conclusion.   

Our paper is organized as follows. In section 2 we present the Lagrangian density and obtain the equations of motion for the vortex. In section 3 we obtain an expression for the ADM mass in an AdS$_3$ and Minkowski background in terms of the asymptotic metrics. We then obtain integral mass formulas over the matter fields for the same backgrounds including the no gravity case (details are relegated to Appendices A and B). In section 4 we present our numerical results in a series of plots and tables of values for both the AdS$_3$ and Minkowski background. We also present some preliminary analytical results. In section 5 we present in a new light the logarithmic divergence in the absence of gauge fields. Our conclusion summarizes our results and discusses future novel directions for this work.

\section{Lagrangian, ansatz and equations of motion for static vortex in Einstein gravity with cosmological constant}
The Lagrangian density for the vortex coupled to Einstein gravity with cosmological constant is given by 
\beq
\mathcal{L} = \sqrt{-g}\Big(\alpha\,(R-2 \Lambda) -\dfrac{1}{4} F_{\mu\nu}F^{\mu\nu} -\dfrac{1}{2}(D_{\mu} \phi)^{\dagger}(D^{\mu} \phi)-\dfrac{\lambda}{4}(|\phi|^2-v^2)^2\Big)
\eeq{LDensity}
where $R$ is the Ricci scalar, $\Lambda$ is the cosmological constant, $F_{\mu\nu}$ is the electromagnetic field tensor and the covariant derivatives are defined in the usual fashion by
\beq
D_{\mu} \phi=\partial_{\mu} \phi +i e A_{\mu} \phi \,.
\eeq{Covariant}
The constant $\alpha$ is equal to $1/(16 \pi G)$ where $G$ is Newton's constant and the constant $v$ is the non-zero VEV of the scalar field which spontaneously breaks the local $U(1)$ symmetry. We consider rotationally symmetric static solutions so that the 2+1-dimensional metric has the form 
\beq
ds^2=- B(r) \,dt^2 +\dfrac{1}{A(r)} \,dr^2 + r^2 d\theta^2\,. 
\eeq{metric}
We make the following ansatz for the scalar and gauge fields 
\beq
\phi(\mathbf{x}) = f(r) e^{i n \theta} \word{and} A_j(\mathbf{x})=\epsilon_{jk}\hat{x}^k \dfrac{a(r)}{er}
\eeq{ansatz}  
where $n$ is the winding number\footnote{The complex scalar field depends on $\theta$ only through its phase so that when it is inserted in the Lagrangian density this yields terms that are dependent on $r$ only. The winding number $n$ appears in the Lagrangian density but not $\theta$. We thus obtain vorticity and equations of motion which are rotationally symmetric.}.  The magnetic field is given by $F_{21}=\dfrac{1}{er} \dfrac{da}{dr}$. Substituting the ansatz \reff{metric} and \reff{ansatz} into the Langrangian density we obtain 
\beq
\mathcal{L}=\sqrt{B/A}\,r\left(\alpha  (R-2\Lambda )-(\lambda /4)( f^2 -v^2)^2 -\frac{(f')^2 A}{2}- \frac{(n-a)^2f^2}{2 r^2}-\frac{A (a')^2}{2 e^2 r^2}\right)
\eeq{LDensity2}
where a prime denotes derivative with respect to $r$. In terms of the metric functions $A$ and $B$ the Ricci scalar is given by
\beq
R= -\frac{ A'}{r }+\frac{ (B')^2 A}{ 2 B^2}-\frac{ B'' A}{B}-\frac{A'B'}{2 B}- \frac{B' A}{r B} \,.
\eeq{Ricci}
The Lagrangian density contains four functions of $r$: $A$, $B$, $f$ and $a$. Their equations of motion are respectively 

\begin{align}
&4 e^2 r \alpha  A B'+ B \Big(e^2 r^2 \left(v^4 \lambda +8 \alpha  \Lambda \right)
+2 e^2 \left(n-a\right)^2 f^2 -2 e^2 r^2 v^2 \lambda f^2 \nonumber\\&\qquad\qquad+ e^2 
r^2 \lambda f^4-2 A \left((a')^2 + e^2 r^2 (f')^2\right)\Big)=0 \label{EOMA}\\\nonumber\\
&2 e^2 \left(n-a\right)^2 f^2 -2 e^2 r^2 v^2 \lambda f^2+e^2 r^2 \lambda  f^4 +e^2 r \left(r v^4 \lambda +8 r \alpha  \Lambda +4 \alpha  A'\right)\nonumber\\&\qquad\qquad+2 A \left((a')^2+e^2 r^2 (f')^2\right)=0\label{EOMB}\\\nonumber\\
&r^2 A B' f'+B \Big(-2 \left(n-a\right)^2 f + 2 r^2 v^2 \lambda f -2 r^2 \lambda  f^3
\nonumber\\&\qquad\qquad + r \left(r A' f'+2 A \left(f'+r f''\right)\right)\Big)=0\label{EOMf}\\\nonumber\\
&r A a^{\prime}B^{\prime} +B \left(2 e^2 r (n-a) f^2-2 A a'+r a' A'+2 r A a''\right)=0\,.\label{EOMa}
\end{align}
We can eliminate $B$ by substituting $B^{\prime}/B$ from equation \reff{EOMA} into equations \reff{EOMf} and \reff{EOMa}. The three final equations of motion that we work with are : 
\begin{align}
&2 e^2 \left(n-a\right)^2f^2 -2 e^2 r^2 v^2 \lambda f^2+e^2 r^2 \lambda  f^4 +e^2 r \left(r v^4 \lambda +8 r \alpha  \Lambda +4 \alpha  A'\right)\nonumber\\&\qquad\qquad+2 A \left((a')^2+e^2 r^2 (f')^2\right)=0\label{EOMB2}\\\nonumber\\
&-2 \left(n-a\right)^2 f + 2 r^2 v^2 \lambda f-2 r^2 \lambda  f^3+\frac{r f^{\prime}}{4 e^2 \alpha }\Big(-e^2 r^2 \left(v^4 \lambda +8 \alpha  \Lambda \right)-2 e^2 \left(n-a\right)^2\,f^2 \nonumber\\ &\quad+2 e^2 r^2 v^2 \lambda f^2-e^2 r^2 \lambda  f^4+2 A \left((a')^2+e^2 r^2 (f')^2\right)\Big)+r \left(r A' f'+2 A \left(f'+r f''\right)\right)=0\label{EOMf2}\\\nonumber\\
&2 e^2 r (n-a) f^2-2 A a'+r a' A'+\frac{a'}{4 e^2 \alpha} \Big(-e^2 r^2 \left(v^4 \lambda +8 \alpha  \Lambda \right)\nonumber\\ &\quad-2 e^2 \left(n-a\right)^2f^2 +2 e^2 r^2 v^2 \lambda f^2-e^2 r^2 \lambda  f^4+2 A \left((a')^2+e^2 r^2 (f')^2\right)\Big)+2 r A a''=0\,.
\label{EOMa2}\end{align}

\subsection{Vacuum and asymptotic metric}
One can solve analytically for the metric in vacuum by setting $f=v$ and $a=n$ identically in Eq. \reff{EOMB2} (this eliminates matter from the Lagrangian density \reff{LDensity2}). This yields $A'(r)=-2 r \Lambda$ with solution
\beq
A_0(r) = -\Lambda r^2 + C
\eeq{A0}
where the subscript `0' denotes vacuum. The integration constant $C$ determines the initial conditions at $r=0$. Besides AdS$_3$ we would also like to study asymptotically Minkowski spacetime ($\Lambda=0$) so in this work we will set $C=1$ (note that this choice also avoids a conical singularity \cite{BTZ1,Deser}). Substituting the above vacuum solution into Eq. \reff{EOMA} yields $B_0(r) = k_0\,(-\Lambda r^2 + C)$ where $k_0$ is a positive integration constant which can be absorbed into a redefinition of time. We therefore obtain 
\beq
B_0(r)= -\Lambda r^2 + C = A_0(r)\,.
\eeq{B0}
In the presence of matter (the vortex), as $r\rightarrow R$, where $R$ is the computational boundary representing formally infinity, we have that $f \rightarrow v$ and $a \rightarrow n$. Using again Eq. \reff{EOMB2}, the asymptotic form of the metric function $A(r)$ in the presence of matter is\footnote{Outside a core radius $r_0$ of the vortex, the mass, the scalar field $f(r)$ and gauge field $a(r)$ hardly change (by less than two percent). Therefore $r_0$ sets the scale by which one can compare the computational boundary $R$. In our plots, a conservative value for the core radius is $r_0=4$ while the computational boundary is $R=10$. So $R=10$ extends well beyond the core of the vortex in the sense that from $r_0=4$ to $R=10$ one is almost entirely in the plateau region where things hardly change. For numerical purposes, we start our plots at $r=0.001$ instead of $r=0$. 
There are three scales in this problem: a UV scale $\ell_{planck}$ and two IR scales, the AdS radius $\ell_{AdS}$ and $(ev)^{-1}$. For the classical approximation to hold, one requires that $\ell_{AdS}$ and $(ev)^{-1} \gg \,\ell_{planck}$. The radius must be expressed in units of the IR scales. Since $-\Lambda\,R^2$ is dimensionless, we can express the radius in units of $\sqrt{-\Lambda} \,\ell_{AdS}$ where $\Lambda$ is a purely negative number e.g. if $\Lambda=-1$ the radius is in units of $\ell_{AdS}$  and if $\Lambda=-3$ the radius is in units of $\sqrt{3}$ $\ell_{AdS}$. One can express $\ell_{AdS}$ in terms of the other IR scale $(ev)^{-1}$. In section 4.1.2 we derive Eq. \reff{Lambda5} which is a relation between the IR scale $\ell_{AdS}$ and $v^{-1}$ i.e. $\ell_{AdS}= (q/e)\,v^{-1}$ where the ratio $q/e$ is of order unity and depends on which of the five AdS cases one is considering. For more details and a definition of the parameter $q$, see sections 4.1.2 and 4.2.} 
\beq
A(R)= -\Lambda R^2 + D  
\eeq{AR}
where the constant $D$ differs from the constant $C$ in \reff{A0}. In the next section we will express the mass (ADM mass) of the vortex in terms of $A_0(R)$ and $A(R)$. 

Using \reff{EOMA} we obtain $B(R)= k \, A(R)$ where the positive integration constant $k$ can no longer be fully absorbed into a redefinition of time (since that has already been done once with the constant $k_0$). Therefore, in the presence of matter, at large radius $R$, we have that $B(R)$ is proportional to $A(R)$ but not equal to it.    

\section{ADM mass and its representation as an integral over matter}   
\subsection{ADM mass} 
The spacetime is asymptotically AdS$_3$, a maximally symmetric spacetime with isometry group $SO(2,2)$. It clearly has a timelike Killing vector since no metric components has a dependence on time. The notion of a conserved energy (the ADM mass) should therefore clearly apply to matter embedded in AdS$_3$. Maximally symmetric spacetimes can be viewed as the ground states of General Relativity \cite{Carroll} so that it is reasonable to set their energy to be zero. This requires one to subtract out from the Hamiltonian the contribution of the background. This subtraction mechanism is already implemented in the usual formulation of the ADM mass because Minkowski spacetime by itself would otherwise contribute a divergent energy \cite{Poisson}. We can therefore calculate the ADM mass using one of its typical forms \cite{Poisson} suitably generalized to $2+1$ dimensions  
\beq
M= -2 \alpha\, \lim_{C_t \to R} \oint_{C_t} (k-k_0) \sqrt{\sigma} N(R) d \theta 
\eeq{M}
where $C_t$ is the circle at spatial infinity (i.e. the computational boundary $r=R$), 
$N(R)= [B_0(R)]^{1/2}=[A_0(R)]^{1/2}$ is the lapse, $\sigma_{AB}$ is the metric on $C_t$, $k$ is the extrinsic curvature of $C_t$ embedded on the two-dimensional spatial surface obtained by setting $t$ to be constant in \reff{metric} and $k_0$ is the extrinsic curvature of $C_t$ embedded in the two-dimensional spatial surface of AdS$_3$. A simple calculation yields 
\beq
k= \dfrac{[A(R)]^{1/2}}{R} \quad;\quad k_0= \dfrac{[A_0(R)]^{1/2}}{R} \quad \word{and}\quad \sqrt{\sigma} =R\,.
\eeq{extrinsic}
The ADM mass is then given by 
\beq
M= 4 \pi \alpha \,\Big(A_0(R) -[A_0(R)A(R)]^{1/2}\Big)\,. 
\eeq{ADMMass2}
As a simple check, if $A(R)=A_0(R)$ then $M=0$ so that empty AdS$_3$ space has zero mass, the expected and desired result. We can simplify the above formula. From Eqs. \reff{A0} and \reff{AR} we have that $A(R)=-\Lambda R^2 +D =A_0(R) + (D-C)$. At large $R$, $A_0(R)>>(D-C)$ so that a binomial expansion of $[A_0(R)A(R)]^{1/2}$ at large $R$ yields $A_0(R) + \dfrac{1}{2} (D-C)$. We therefore obtain that $M= 2 \pi \alpha \,(C-D)$. The final expression for the mass in an AdS$_3$ background is then\footnote{Eq. \ref{M2} is valid when $|A(R)/A_0(R) -1|<<1$. This is completely satisfied in all the cases we ran (see values in Table 1). For example, in case I of Table 1, $|A(10)/A_0(10) -1|=0.0004$ and the percentage difference between the mass given by Eq. \ref{ADMMass2} and Eq. \ref{M2} is $0.0008\%$. This is such a tiny difference that for all intents and purposes it can be neglected. The advantage of using Eq. \ref{M2} over Eq. \ref{ADMMass2} is that it leads to a more compact and less cumbersome expression for the mass integration formula.}  
\beq
M_{AdS_3}= 2 \pi \alpha \,(A_0(R)-A(R))\,.
\eeq{M2}  
Note that we recover the BTZ black hole spacetime asymptotically. In the BTZ literature \cite{BTZ1,BTZ2}, $\alpha$ is set to $1/(2 \pi)$ and $C$ to zero. We then obtain $D=-M$ so that $A(R)=-\Lambda R^2 -M$ and $B(R) =k\,A(R)$ which represents the BTZ black hole spacetime asymptotically (the presence of the positive constant $k$ in $B(R)$ has no effect on the spacetime: all curvature scalars are independent of $k$).  

The value of $A_0(R)=-\Lambda R^2 +C$ can be calculated analytically for a given $R$. As mentioned earlier, in our work we set $C=1=A_0(0)$. The value of $A(r)$ at $r=0$ is not determined by the equations of motion; it is an initial condition. Since $A(r)$ must reduce to $A_0(r)$ in the absence of matter, their values must match at the origin. We therefore have $A(0)=A_0(0)=C=1$. The quantity $A(R)$ is then obtained by solving the equations of motion numerically and the mass $M$ obtained via Eq.\reff{M2}. One has reached a large enough value of $R$ when the matter functions $f(r)$ and $a(r)$ reach a nice plateau at the values of $v$ and $n$ respectively and the mass $M$ is stable i.e. that in reducing 
$R$ in the plateau region, $A_0(R)$ and $A(R)$ change but not the value of $M$.

The ADM mass formula \reff{ADMMass2} applies to asymptotically flat spacetime where $\Lambda=0$. In that case we obtain $A_0(R)=C$ and $A(R)=D$ so that
\beq
M_{flat}= 4 \pi \alpha \,\Big(C -(C\,D)^{1/2}\Big) =4 \pi \alpha \,\Big(1-D^{1/2}\Big)
\eeq{MFlat} 
where we used $C=1$. In asymptotically flat spacetime, $A_0(r)=B_0(r)=1$ (stays constant for all $r$) whereas $A(r)$ starts at unity at $r=0$ and then decreases with $r$ until it reaches a plateau at a positive value of $D$ that is less than unity. The value of $D$ is obtained numerically. Note that formula \reff{MFlat} is in agreement with the angular deficit expected to be produced by localized sources in a $2+1$ dimensional asymptotically flat spacetime\footnote{In source free regions, the Einstein equations in $2+1$ dimensions with no cosmological constant yields a locally flat spacetime. Outside localized sources, the spacetime is therefore flat but the topology is that of a cone. This can be seen by rewriting the metric in the source free region in a manifestly flat form and observing that the new angle no longer ranges from $0$ to $2 \pi$ but covers a smaller range so that there is an angular deficit. The angular deficit is proportional to the mass of the localized source.} \cite{Deser}. Replacing $\alpha$ by $1/(16 \pi G)$, formula \reff{MFlat} reads $D^{1/2}= 1- 4 G M_{flat}$ which is nothing other than the parameter ``$\alpha$" (not related to our $\alpha$) found in \cite{Deser} that determines the angular deficit.  If we work in new coordinates $(r',\theta')$ where $r'=r/D^{1/2}$ and $\theta'= D^{1/2} \theta$ the two-metric asymptotically takes the flat form $dr'^2 + r'^2 \,d\theta'^2$ with $\theta'$ ranging from $0$ to $2 \pi D^{1/2}$ so that there is an angular deficit of 
\beq
\delta= 2 \pi (1-D^{1/2})\,.
\eeq{delta}
The mass with no gravity in asymptotically flat spacetime, $M_{no-gravity}$, cannot be obtained from comparing two metrics because in this case the metric is fixed throughout to be Minkowski spacetime. However, we derive an integral mass formula for $M_{no-gravity}$ in Appendix A by taking the limit as $\alpha \to \infty$ of $M_{flat}$. The results are reported in the next section.  

\subsection{Integral mass formulas}
The second and third equations of motion \reff{EOMf2} and \reff{EOMa2} can be solved for $A(r)$ in terms of matter fields. This can then be substituted into the first equation \reff{EOMB2} to obtain $A'(r)$ in term of matter only (see Appendix A). One then obtains the following integral representation for the ADM mass \reff{M2} of a vortex embedded in an AdS$3$ background  
\beq 
M_{AdS_3}=I
\eeq{M4}
where $I$ is given by \reff{Int2}
\begin{align} 
\label{Int3}
I&= \frac{\pi}{2 e^2}\int_0^R \frac{1}{r}\Bigg[e^2 r^2 v^4 \lambda +2 e^2 \left(n^2-r^2 v^2 \lambda -2 n a+a^2\right) f^2+e^2 r^2 \lambda  f^4\\\nonumber&\quad\quad\quad+\dfrac{2}{r \left(2 a' f'-r f' a''+r a' f''\right)}\Big(n^2 f a'-r^2 v^2 \lambda  f a'-2 n a f a'+a^2 f a'\\\nonumber&\qquad\qquad\qquad+r^2 \lambda  f^3 a' +e^2 n r^2 f^2 f'-e^2 r^2 a f^2 f'\Big) \left(a'^2+e^2 r^2 f'^2\right)\Bigg]\, dr\,.
\end{align}
Note that $I$ does not make any reference to metrics, Newton's constant or the cosmological constant\footnote{The function $f(r)$ has dimension of (length)$^{-1/2}$ and so does the VEV $v$. The function $a(r)$ is dimensionless. Therefore the integral $I$ has dimensions of (length)$^{-1}$ which is the dimension of mass.}. It is an integral that is purely over the matter field profiles. 

For a vortex under gravity in an asymptotically flat spacetime the integral  representation for the ADM mass \reff{MFlat} is given by \reff{MFlat2} 
\beq
M_{flat}= 4 \pi\alpha \Big(1- \sqrt{1-\chi}\Big)
\eeq{MFlat3}
where 
\begin{align} 
\chi=\dfrac{I}{2\pi\alpha}\,.
\end{align}
$I$ is the integral given by \reff{Int3}. So $\chi$ is an integral over matter field profiles just like $I$ except that it has a dependence on $\alpha$.     

Solving the equations of motion numerically yields the metric and matter field profiles. We can evaluate $M_{AdS_3}$ and $M_{flat}$ in two ways: using the metrics or using the matter. In the first method, we extract the asymptotic value of the metric in vacuum and the metric with matter and use the ADM mass formulas \reff{M2} and \reff{MFlat} respectively. In the second method one substitutes the matter field profiles into the integral mass formulas \reff{M4} and \reff{MFlat3} respectively. The two methods must match and this provides a good check on our numerical results. 

The first method of extracting values from two separate metrics does not work for the case with no gravity because it takes place in a fixed Minkowski spacetime. However, by taking the limit as $\alpha \to \infty$ of $M_{flat}$ one can derive an integral mass formula that works perfectly well in this case (see Appendix A). The mass of the vortex without gravity given by \reff{NoGravity} is 
\beq
M_{no-gravity} =I
\eeq{NoGravity2}
where $I$ given by \reff{Int3} is the same integral that we already encountered for $M_{AdS_3}$. Even though we evaluate $M_{AdS_3}$ and $M_{no-gravity}$ using the same integral mass formula, the matter field profiles for a vortex under gravity in an AdS$_3$background will clearly differ from those of a vortex with no gravity in a fixed Minkowski spacetime, leading naturally to different masses. 

It is worth noting that in texts that discuss the vortex in fixed Minkowski spacetime (no gravity) they derive from the Hamiltonian an integral expression for the energy of the vortex (label it $I_{no-gravity}$). However, its integrand looks quite different from that of $I$ given by \reff{Int3}. Of course, the two formulas should yield the same mass for the no gravity case and they do. In Appendix B, \textit{using the equations of motion for the no gravity case}, we show that the integral $I$ can be transformed into the integral $I_{no-gravity}$. It is important to note that $I$ and $I_{no-gravity}$ are not equal at a purely mathematical level. They can be shown to be equal only when the equations of motion for the no gravity case are used. In particular this means that $I$ (and not $I_{no-gravity}$) should be used to calculate the mass for the case of an AdS$_3$ background.

We can rewrite the integral $I$ given by \reff{Int3} in a convenient form that reveals more clearly its dependence on the VEV of the scalar field $v$ and the winding number $n$. Since $f(r)$ and $a(r)$ plateau at $v$ and $n$ respectively we can write $f(r)=v\,f_1(r)$ and $a(r)=n\, a_1(r)$ where $f_1(r)$ and $a_1(r)$ will always reach asymptotic values of (plateau at) unity. Define $u=\tfrac{e\,v}{n}\,r$. Then $a'(r)= e\,v a_1'(u)$, $a''(r)=
\tfrac{e^2\,v^2}{n} a_1''(u)$, $f'(r)=\tfrac{e\,v^2}{n}\,f_1'(u)$, $f''(r)=\tfrac{e^2\,v^3}{n^2} \,f_1''(u)$ and $dr=n\,du/(ev)$. Derivatives on functions with subscript `1' are with respect to $u$. Substituting this into \reff{Int3} yields\footnote{The integral $F$ is dimensionless and the dimension of mass stems from $v^2$ which has dimensions of (length)$^{-1}$. Note that $\lambda/e^2$ is dimensionless.} 
\beq
 I= n^2 v^2 F\big(\frac{\lambda}{e^2}\big)
\eeq{I3}
where the integral $F$ is given by
\begin{align}
&F\big(\frac{\lambda}{e^2}\big)\\
&=\dfrac{\pi}{2}\int_0^{R_1} \frac{1}{u}\Bigg[u^2 \frac{\lambda}{e^2} + f_1^4 u^2\frac{\lambda}{e^2}  + 2 f_1^2 \big((-1 + a_1)^2 - u^2 \frac{\lambda}{e^2} \big) \nonumber\\& \quad+ \frac{2 f_1 \big((a_1^{\prime})^2 + (f_1^{\prime})^2 u^2\big) \Big((1 - a_1) f_1 f_1^{\prime} u^2 + a_1^{\prime} \big((-1 + a_1)^2  + (-1 + f_1^2) u^2 \frac{\lambda}{e^2}\big)\Big)}{u (2 a_1^{\prime} f_1^{\prime} - a_1^{\prime\prime} f_1^{\prime} u + a_1^{\prime} f_1^{\prime\prime} u)}\Bigg] du\,.
\label{F3}
\end{align}  
Here $R_1=\frac{e\,v}{n} R$ and $F$ is a function of $\lambda/e^2$ and a functional of the matter profiles $f_1(u)$ and $a_1(u)$.  The formula \reff{I3} does not necessarily imply that $I$ grows exactly quadratically with $v$ and $n$ because the matter profiles that appear in $F$ change with $v$ and $n$. However, since $f_1(u)$ and $a_1(u)$ always plateau to unity, their change in profile is expected to be limited so that $I$ should still increase significantly if, for example, we double $v$ or $n$. This is what is observed numerically.

\section{Numerical results}

We solve the three equations of motion \reff{EOMB2}, \reff{EOMf2} and \reff{EOMa2} numerically for the non-singular profiles of the scalar field $f(r)$, the gauge field $a(r)$ and the metric $A(r)$ using a negative cosmological constant $\Lambda$ (AdS$_3$ background). We have the following boundary conditions:
\beq
f(0)=0\quad; \quad a(0)=0\quad;\quad f(R)=v\quad;\quad a(R)=n\quad;\quad A(0)=1
\eeq{boundary}
where $R$ is the computational boundary representing formally infinity. The quantity $v$ is the VEV of the scalar field and $n$ is the winding number of the vortex. $A(0)=1$ is an initial condition at $r=0$ that is in accord with setting $C=1$ in the vacuum solution $A_0(r)$ given by \reff{A0}. We obtain the profiles by adjusting $f'(r)$ and $a'(r)$ near the origin to give the final boundary conditions at $R$ where both $f$ and $a$ plateau to their respective values. 

\subsection{Preliminary analytical results}

Before presenting the numerical results, it is worthwhile to extract a few analytical results from the equations of motion.

\subsubsection{Analytical behaviour of $f(r)$ and $a(r)$ near the origin}
We require that at the origin $r=0$, $f(0)=0$ and $a(0)=0$. If we linearize equation \reff{EOMf2} about $f=0$ we obtain that near the origin $f(r)=b\,r^n$ where $n$ is the winding number and $b$ is a positive constant.  This implies that near the origin $f'(r)=b\,n r^{n-1}$ so that for $n=1$, $f'(0)=b$ and for $n>1$, $f'(0)=0$. So for $n=1$, $f(r)$ has a positive slope at $r=0$ whereas it starts off flat for any higher winding number. This is what is observed numerically.  If we linearize \reff{EOMa2} about $a=0$ we obtain $a(r)=c\,r^2$ where $c$ is some positive constant. This implies that $a'(0)=0$ so that the profile always starts out flat. Again, this is what is observed numerically.   

\subsubsection{Analytical expression for the cosmological constant} 
The matter fields $a$ and $f$ reach asymptotic values of $n$ and $v$ respectively. At large $r$ we can write $a(r)=n-\epsilon(r)$ and $f(r)=v-\sigma(r)$ where $\epsilon$ and $\sigma$ are small perturbations that approach zero asymptotically. Substituting these expressions into equation \reff{EOMa2} and keeping only terms linear in $\epsilon$ and $\sigma$ yields the following differential equation: $e^2 v^2 \epsilon +\Lambda r \epsilon' +\Lambda r^2 \epsilon''=0$ where $\Lambda$ has been assumed to be non-zero. This has a power law fall off solution $\epsilon=b/r^q$ where $b$ and $q$ are positive constants. Substituting this solution into the differential equation yields $ e^2 v^2-\Lambda q+\Lambda q(q+1)=0$ with solution  
\beq
\Lambda= -\dfrac{e^2\,v^2}{q^2}
\eeq{Lambda5}
an analytical expression for the cosmological constant in terms of the power $q$. The important thing here is that $\Lambda$ is negative; this implies that in $2+1$ dimensional Einstein gravity there are no de Sitter vortices just like there are no de Sitter black holes. In contrast, de Sitter Einstein-Yang-Mills-Higgs magnetic monopoles exist in $3+1$ dimensions \cite{AE}. The above derivation assumed $\Lambda \ne 0$. There are of course vortices in asymptotically Minkowski spacetime ($\Lambda=0$) which we consider later\footnote{Asymptotically Minkowski spacetime requires that the cosmological constant $\Lambda$ be identically zero. There is a discontinuity between the limit as $\Lambda$ tends to zero and $\Lambda=0$ \cite{Ashtekar1,Ashtekar2}. This means that equation \reff{Lambda5}, which is valid only for $\Lambda \ne 0$, cannot be used to obtain the asymptotically flat limit.}. 

When we solve the equations of motion numerically, the parameters $e$, $v$ and $\Lambda$ are given. Equation \reff{Lambda5} can then be used to solve for $q$ to predict how $a$ approaches $n$ asymptotically:
\beq
q=\dfrac{e\,v}{(-\Lambda)^{1/2}}\,.
\eeq{q2}
We can verify the above equation with our numerical results (this is done near the end of the next subsection.) 

 \subsection{Numerical results for AdS$_3$ background}

For an AdS$_3$ background, there are six parameters in our numerical simulations: 
$\lambda$, $\alpha$, $e$, $n$, $v$ and $\Lambda$. We set $\lambda=1$, $\alpha=1$ and $e=3$. We ran numerical simulations for five different cases determined by the values of the three parameters ($n$, $v$, $\Lambda$). The five cases are I=(1,1,-1),II= (1,1,-2),III=(1,1,-3), IV=(1,2,-2) and V=(2,1,-2). 

We calculate the mass of the vortex in two ways: first, using the metric together with Eq. \reff{M2} and second, inserting the matter profiles into integral \reff{M4}. The masses are listed in Table 1 and labelled $M_{metric}$ and $M_{integral}$ respectively. The two values match to within two or three decimal places. We know we have reached a large enough $R$ value when the mass is stable (i.e. when we take values of the functions in the plateau region at a radius $r$ much less than $R$ and the mass remains the same). In all cases $R=10$ sufficed\footnote{  In our plots, the computational boundary was $R=10$ and at that radius the matter functions $f(r)$ and $a(r)$ reached $99.9 \%$ of their expected asymptotic values. A computational limitation of our procedure is that our simulation starts at a radius of $r=0.001$ instead of $r=0$. So strictly speaking, we could not implement exactly the boundary conditions at $r=0$ like $f(0)=0$ and $a(0)=0$. However, as mentioned above, the matter functions did reach their expected asymptotic values to within $99.9 \%$ so this limitation does not seem to have affected the end results of the simulation in any significant fashion.}.

For each of the five cases we plot in figures 1 to 5 below the following functions: the matter functions $f$ and $a$, the metric functions $A$ and $B$ and the 
Ricci scalar $R$. In all cases, the metric functions $A(r)$ and $B(r)$ are positive throughout so that there is no event horizon. From the Einstein field equations in $2+1$ dimensions, we expect $R$ to plateau at its vacuum value of $6\Lambda$ which is what is observed. We also plot the metric function $A$ near the origin. This is to demonstrate that near the origin and hence in the core of the vortex, $A$ will depart from its ``asymptotic" $r^2$ dependence. This is particularly pronounced in cases $IV$ and $V$ where $A$ dips significantly below unity near the origin while remaining positive. For these two cases, near the core of the vortex, the metric looks nothing like the BTZ black hole metric. The caption for each of the five figures contains pertinent information besides what we mention below in the text. 

Cases $I$, $II$ and $III$ have the same value of $n=1$ and $v=1$ but differ in their value of $\Lambda$ which are $-1$, $-2$ and $-3$ respectively. In Table 1, the mass increases as one goes from case $I$ to case $III$. The vortex is also more compressed when $\Lambda$ is more negative; the value of $r$ at which $f(r)$ reaches $0.9$ are $1.07$, $0.92$ and $0.83$ respectively so that the core of the vortex gets progressively smaller as $\Lambda$ becomes more negative. When matter is compressed it usually gains positive energy so that it makes sense that the mass has increased.

The mass $M=6.53$ of case $IV$ which has $v=2$ and the mass $M=7.14$ of case $V$ with $n=2$ is significantly greater than the mass $M=2.93$ of case $II$ (which has two parameters in common with case $IV$ and case $V$ separately). This is in accord with the $n^2v^2$ coefficient in the integral mass formula \reff{I3}. Here case $IV$ and case $V$ are $2.22$ and $2.44$ times more massive than case $II$. As already pointed out at the end of section $3.2$, it is not four times greater because the matter profiles $f_1$ and $a_1$ that enter \reff{I3} change with $v$ and $n$. 

We now check how well equation \reff{q2} matches the numerical results. We label the $q$ appearing in \reff{q2} as $q_{predicted}$. For cases $I$ to $V$ we obtain (quoting numbers sequentially) $q_{predicted}= [3, 2.12,1.73,4.24,2.12]$. By taking points asymptotically from the profile of $a$ for each of the five cases and assuming the power law fall off $\epsilon=b/r^q$, we can extract $q_{numerical}$. The values are $q_{numerical}=[2.84,2.03,1.68,4.11,2.06]$. We can see that the numbers match quite well (all less than $5.7 \%$ difference) so that formula \reff{q2} is a reliable predictor of how $a$ behaves asymptotically.                     

\begin{figure}[!htb]
	\centering
		\includegraphics[scale=1.0]{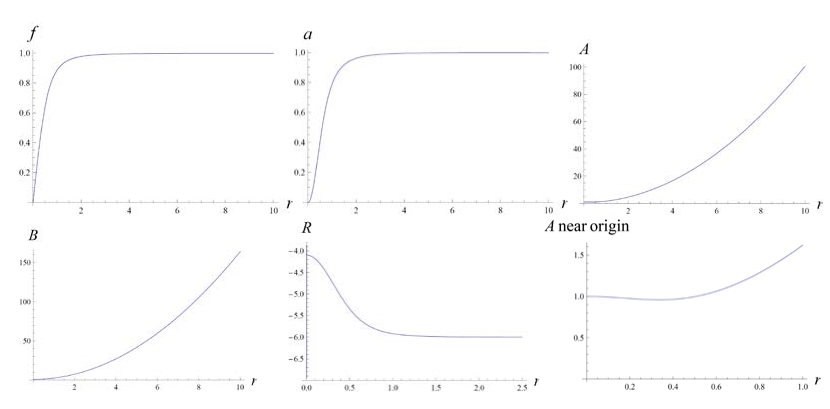}
		\caption{Case I: $n=1,v=1$ and $\Lambda=-1$. The matter functions $f$ and $a$ plateau at their respective values of $v=1$ and $n=1$. The metric functions $A$ and $B$ are positive throughout and hence there is no event horizon (this fact will be true of the next four figures so we will not repeat it later). The Ricci scalar plateaus at $-6$ which agrees with the expected $6\Lambda$. The numerical value of the metric function $A$ at $R=10$ is $100.588$ while the value of $A_0$ at $R=10$ can be evaluated analytically via $-\Lambda r^2+1$ and yields $101$. The mass obtained from \reff{M2} is then $M= 2 \pi \alpha (101-100.588)=2.589$ where $\alpha=1$ was used. The integral mass formula \reff{M4} for the matter profiles plotted here yields $M=2.587$. The two masses agree to within two decimal places. We will not repeat the mass calculation in the captions of the next four cases as it is a similar calculation. The mass results of all five cases are summarized in Table 1.}     
	\label{Graph1}
\end{figure}
\clearpage
\begin{figure}[!htb]
	\centering
		\includegraphics[scale=1.0]{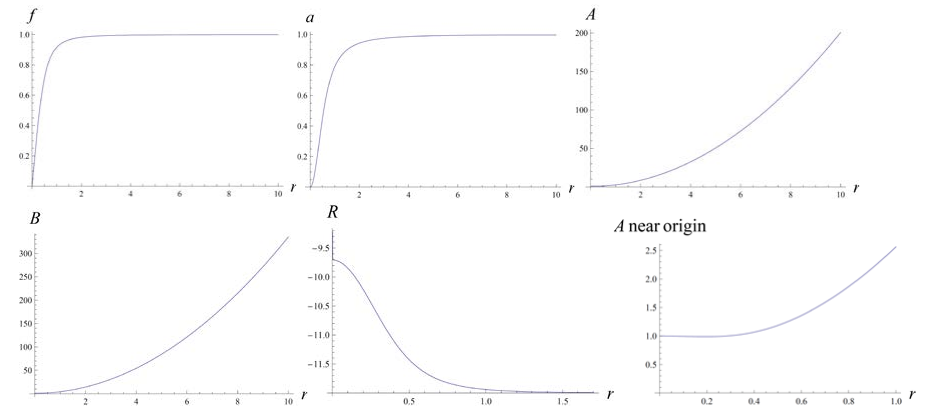}
		\caption{Case II: $n=1,v=1$ and $\Lambda=-2$. The Ricci scalar plateaus at $-12$ which agrees with the expected $6\Lambda$. The core of the matter profile $f$ is smaller (the vortex is more compressed) than in case I and its mass is greater.}     
	\label{Graph2}
\end{figure}

\begin{figure}[!htb]
	\centering
		\includegraphics[scale=1.0]{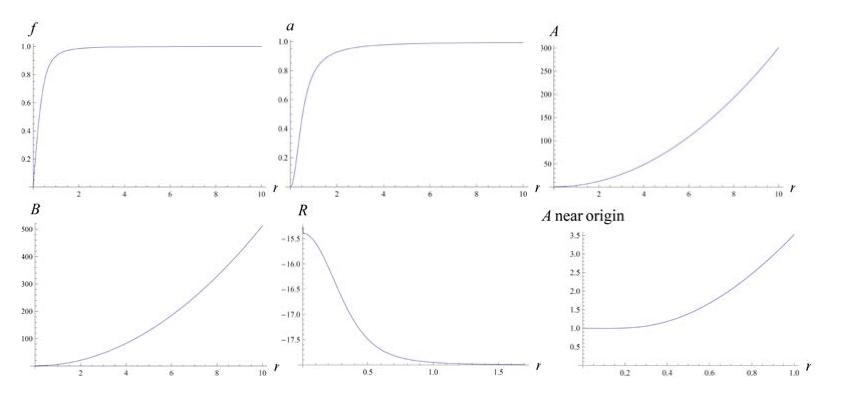}
		\caption{Case III: $n=1,v=1$ and $\Lambda=-3$. The Ricci scalar plateaus at $-18$ in agreement with the value $6\Lambda$. The matter profile $f$ here is more compressed (the core is smaller) than in the two previous cases and it has the greatest mass of the three.}       
	\label{Graph3}
\end{figure}

\begin{figure}[!htb]
	\centering
		\includegraphics[scale=1.0]{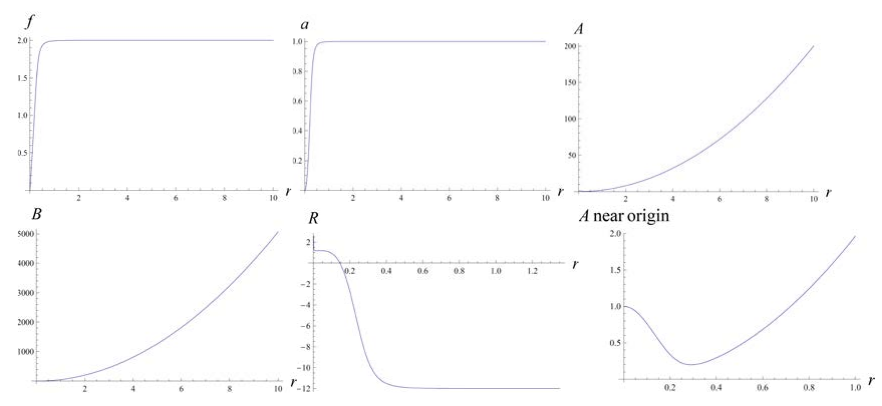}
		\caption{Case IV: $n=1,v=2$ and $\Lambda=-2$. This case differs from the previous three because $v=2$ instead of $v=1$ (hence $f$ plateaus at $2$). The mass of the vortex is considerably greater now compared to the previous three cases (see Table 1 of values) which reflects the $v^2$ dependence of the integral mass formula \reff{I3}. There is a significant dip in the metric $A$ near the origin while it remains positive (no horizon). This reflects a significant departure of the metric near the vortex from its asymptotic BTZ black hole form (which has an $r^2$ dependence).}
	\label{Graph4}
\end{figure}

\begin{figure}[!htb]
	\centering
		\includegraphics[scale=0.9]{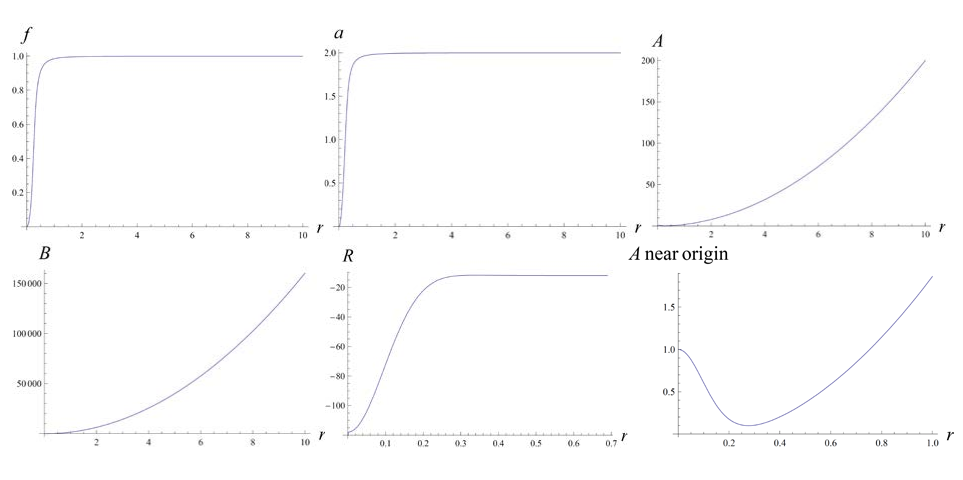}
		\caption{Case V: $n=2,v=1$ and $\Lambda=-2$. This case differs from all four previous cases because the winding number is now $n=2$ instead of $n=1$ (the function $a$ plateaus at $2$ now). It is the most massive case (considerably more than the first three cases and still greater than the $v=2$ case). This reflects the $n^2$ dependence of the integral mass formula \reff{I3}. Again, there is a significant dip in the metric $A$ near the origin while it remains positive (no horizon). Near the core of vortex, the metric departs significantly from its asymptotic BTZ black hole form.}     
	\label{Graph5}
\end{figure}

\begin{table}[!htb]
	\centering
		\includegraphics[scale=0.9]{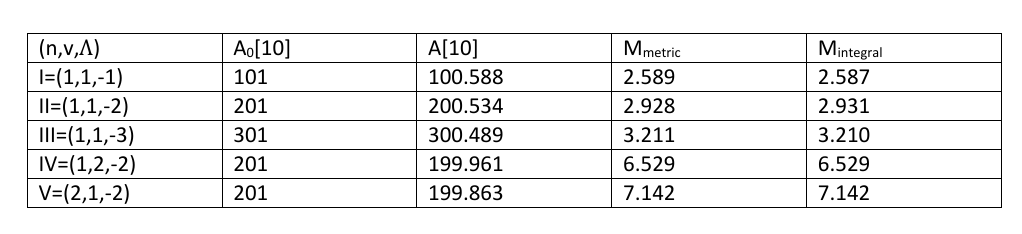}
		\caption{Table with values of the metric $A_0$ and $A$ at $r=10$, $M_{metric}$ evaluated using $A_0$ and $A$ in Eq. \reff{M2} and $M_{integral}$ evaluated using the integral mass formula \reff{M4}. The two masses match (agree to two decimal places and sometimes at three decimal places).}     
	\label{Table1}
\end{table}
\FloatBarrier
\subsection{Numerical results for Minkowski background}

In this subsection we consider vortices under Einstein gravity in $2+1$ dimensions in an asymptotically Minkowski spacetime ($\Lambda=0$). Besides the standard vortex without gravity (fixed Minkowski spacetime) we considered three cases under gravity with different parameters $\alpha=1/(16 \pi G)$. 

In the no gravity case, where the metric is fixed, we only have the profiles of $f$ and $a$ to plot. Even though its metric is fixed, via a limiting procedure, we were able to obtain the integral mass formula \reff{NoGravity2} where $I$ is given by \reff{Int3}. The equations of motion are (it is worth rewriting them for the no gravity case as they simplify considerably)
\begin{align}
&r f \left(-\frac{(n-a)^2}{r^2}+\lambda  \left(v^2-f^2\right)\right)+f'+r f''=0\\
&e^2 r (n-a) f^2-a'+r a''=0\,.
\end{align}
We solve the above equations numerically with the same boundary conditions as \reff{boundary} (except that $A(0)=1$ is not used since the metric function $A(r)$ does not appear in the equations of motion). The plots (figure 6) are below (the parameters used are $e=1$, $\lambda=1$, $v=1$ and $n=1$). The mass evaluated by inserting the profiles of $f$ and $a$ below into \reff{NoGravity2} yields $M_{no-gravity}=3.634$. Inserting the same profiles into the standard integral $I_{no-gravity}$ given by \reff{INoGrav} yields exactly the same value of $3.634$.

For the case of the vortex under gravity in a Minkowski background we solve the equations of motion \reff{EOMB2}, \reff{EOMf2} and \reff{EOMa2} numerically with the boundary conditions \reff{boundary}. The parameters used are: $e=1$, $\lambda=1$, $v=1$, $n=1$ and $\Lambda=0$. We considered three separate cases: $\alpha=1,\alpha=5$ and $\alpha=10$. For each of the three cases we plot the metric function $A$, the Ricci scalar $R$, the metric function $B$ and the matter functions $f$ and $a$. These appear in figures 7, 8 and 9 respectively. Table 2 contains the values of $D$ where the metric function $A$ plateaus, the angular deficit $\delta$ calculated using \reff{delta} and converted in degrees, and the masses of the vortex calculated using the metric (Eq. \reff{MFlat}) and the matter profiles (Eq. \reff{MFlat3}. The mass of the vortex hardly changes with $\alpha$ and hardly differs from the case with no gravity. However, the value of $D$ changes with $\alpha$. This results in an asymptotic conical spacetime where the angular deficit $\delta$ increases significantly as $\alpha$ decreases. There is no singularity at the origin and the Ricci scalar reaches its highest value there before it decreases asymptotically to zero. The value of the Ricci scalar at the origin changes significantly with $\alpha$: it is highest at $\alpha=1$ (R=1.065) and lowest at $\alpha=10$ (R=0.115). We therefore have a smooth spacetime that transitions from a region with curvature at the origin and near the core of the vortex to an asymptotically flat conical spacetime with angular deficit. The curvature at the origin and the angular deficit are highest at $\alpha=1$ and basically five and ten times smaller at $\alpha=5$ and $\alpha=10$ respectively.      

\begin{figure}[!htb]
	\centering
		\includegraphics[scale=0.95]{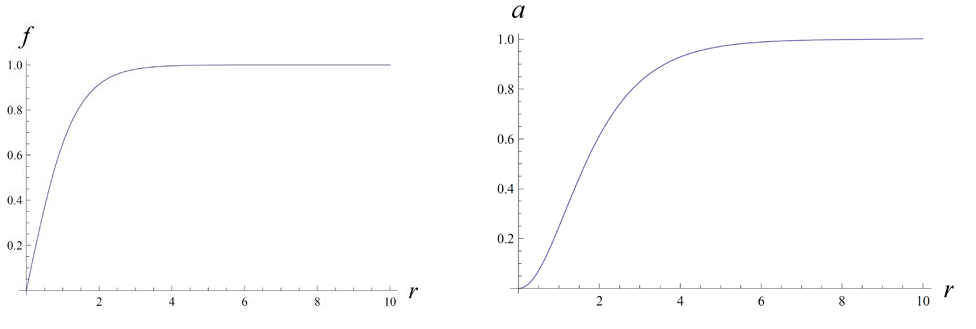}
		\caption{Vortex with no gravity in a fixed Minkowski spacetime. The profiles of $f$ and $a$ are used to evaluate the mass.}
	\label{GraphNoGravity}
\end{figure}   

\begin{figure}[!htb]
	\centering
		\includegraphics[scale=0.9]{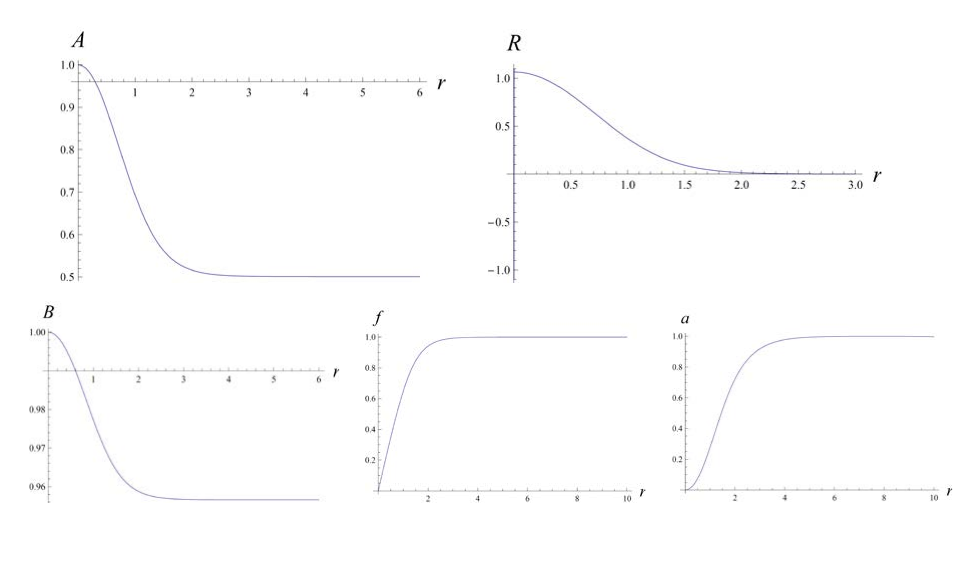}
		\caption{Case $\alpha=1$ in asymptotically Minkowski spacetime. This is the case with the strongest gravitational coupling of the three. The Ricci scalar has the highest initial value of the three cases and the metric function $A$ plateaus to the lowest value of $D$ leading to the highest angular deficit. The mass is not too different from the other two cases or from the no gravity case. See Table 2 for values.}
	\label{Graph_alpha1_flat}
\end{figure} 
\begin{figure}[!htb]
	\centering
		\includegraphics[scale=1.0]{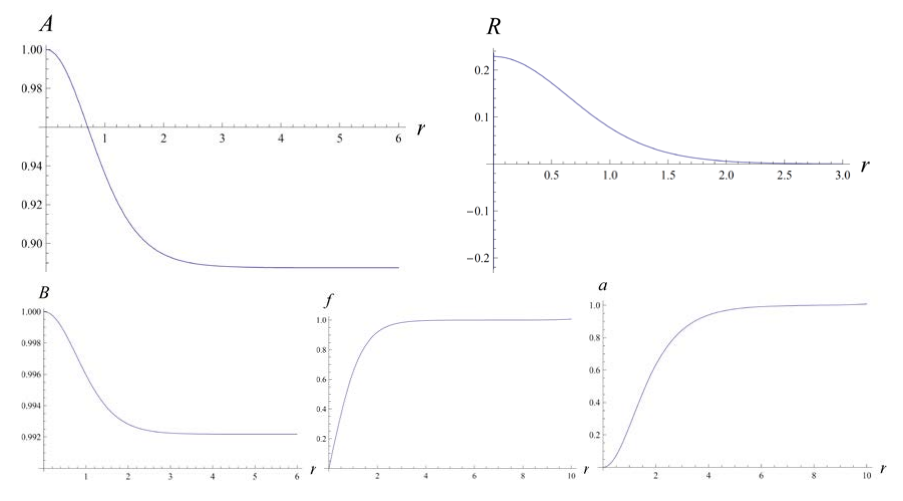}
		\caption{Case $\alpha=5$. The Ricci scalar at the origin and the angular deficit is basically five times less than in the $\alpha=1$ case. The mass is comparable to the other cases.}
	\label{Graph_alpha5_flat}
\end{figure} 

\begin{figure}[!htb]
	\centering
		\includegraphics[scale=1.0]{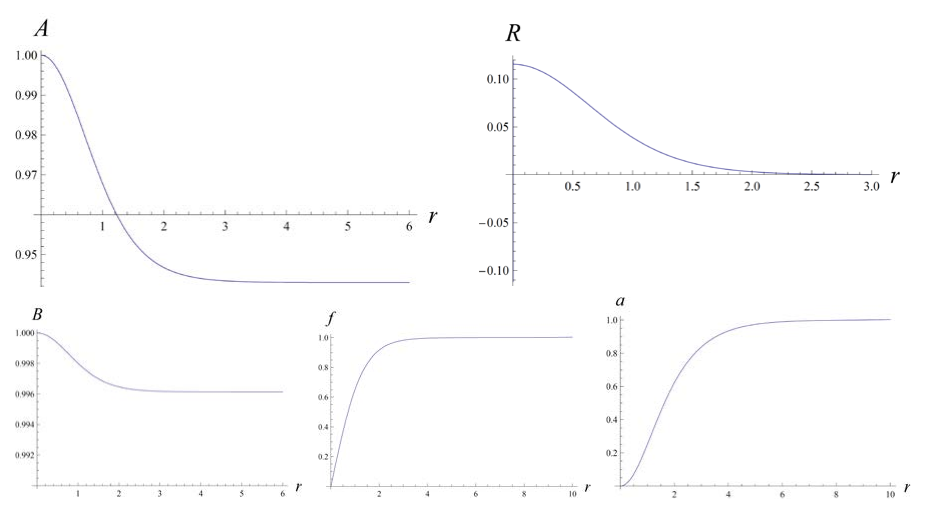}
		\caption{Case $\alpha=10$. This is the case with the weakest gravitational coupling where the initial value of the Ricci scalar and the angular deficit are the smallest (basically 10 times smaller than in the $\alpha=1$ case). The mass is again comparable to the others.}
	\label{Graph_alpha10_flat}
\end{figure} 

\begin{table}[!htb]
	\centering
		\includegraphics[scale=0.9]{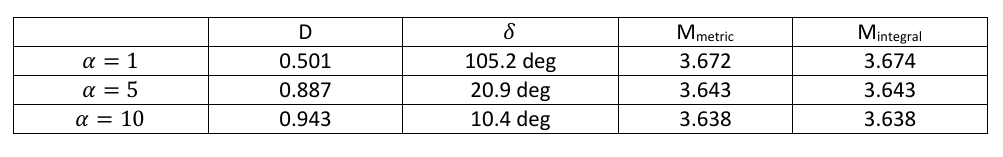}
		\caption{The quantity $D$ is where the metric function $A(r)$ plateaus to asymptotically and $\delta$ is the angular deficit quoted in degrees and evaluated using \reff{delta}. The mass $M_{metric}$ is obtained using the metric and calculated via \reff{MFlat} and $M_{integral}$ is evaluated via \reff{MFlat3} as an integral over the matter profiles. The two masses should match and they do. The angular deficit has a strong dependence on $\alpha$: it basically increases tenfold from $\alpha=10$ to $\alpha=1$. In contrast, the mass of the vortex hardly changes with $\alpha$ and is not very different from the mass in the no gravity case.}     
	\label{Table2}
\end{table}

\section{The logarithmic divergence in the absence of gauge fields: a new look}

Without gauge fields, it is well known that the vortex has a logarithmic divergence in its energy \cite{Weinberg}. This problem does not go away in the presence of gravity.  We will show this by looking at the form of the metric itself. We will see that the metric acquires a logarithmic term which directly implies that the ADM mass diverges. However, the logarithmic term can now be viewed as the realization of the Newtonian gravitational potential in $2+1$ dimensions. 

If there are no gauge fields, we set $a=0$ identically. The vacuum metric is then obtained by setting $f=v$ identically and $n=0$ in equation \reff{EOMB2}. This yields
$A'(r)=-2 \Lambda r$ with solution
\beq
A_0(r)=-\Lambda r^2 +C
\eeq{vaccum3}
which is the same vacuum metric we had obtained before with $C$ an integration constant. Now consider the case where $n\ne 0$, $a=0$ identically and $f \to v$ asymptotically. Substituting this into \reff{EOMB2} yields asymptotically $A'(R)=-2 \Lambda R -\tfrac{n^2\,v^2}{2 \alpha R}$ with solution
\beq
A(R) =-\Lambda R^2 -\dfrac{n^2 v^2}{2 \alpha} \ln(R) + D
\eeq{A4}
where $D$ is an integration constant. The logarithmic term in the metric implies immediately that the energy (the ADM mass) diverges logarithmically.

Replacing $\alpha$ by $1/(16 \pi G)$ the logarithmic term takes the form $ G \,m\,\ln(R)$ with $m=8 \pi n^2 v^2$. This is nothing other than the Newtonian gravitational potential in $2+1$ dimensions with mass parameter $m$ proportional to $n^2v^2$, the same mass dependence we encountered previously. It would be interesting to know whether one can generate a finite energy vortex with logarithmic Newtonian potential when General Relativity is supplemented with a scalar field and appropriate potential. We discuss this further in the conclusion.  

\section{Conclusion}

In this work, we obtained numerical solutions of non-singular vortices of positive mass in $2+1$ dimensional Einstein gravity in both an AdS$_3$ and Minkowski background. We obtained the scalar and gauge matter profiles as well as the metric profiles for different cosmological constant $\Lambda$, winding number $n$ and VEV $v$. The metric is always positive and there is no event horizon, so that in contrast to previous work \cite{Cadoni}, these are not black hole vortex solutions. We derived two different ways to calculate the mass: one using the metrics and one using an integral over the matter profiles. We observed that the vortices increased in mass and became more compressed (core radius smaller) as the cosmological constant became more negative. This makes sense since an object which becomes compressed usually gains positive energy. The cases with $n=2$ and $v=2$ differed from their $n=1$ and $v=1$ counterparts in two regards. First, they had significantly higher masses reflecting the quasi-dependence of the integral mass formula on $n^2 \,v^2$ (``quasi" because as explained at the end of section $3.2$ this dependence is not exact). Secondly, the metric function $A(r)$ near the origin had a significant dip below unity (while remaining positive) with a significant departure from its asymptotic $r^2$ dependence. In other words, near the origin and hence the core of the vortex, the metric looks nothing like the BTZ black hole metric.  

We showed that the cosmological constant must be negative or zero, which implies that de Sitter vortices, just like de Sitter black holes, do not appear to exist in $2+1$ dimensional Einstein gravity. Note that de Sitter black holes exist in BHT massive gravity \cite{BHT} which also takes place in $2+1$ dimensions. This hints at the possibility that BHT massive gravity might support de Sitter non-singular vortices, something that is worth investigating. 

We then considered vortices under gravity in asymptotically Minkowski spacetime ($\Lambda=0$) for different values of the parameter $\alpha$. This leads to an asymptotic conical spacetime with angular deficit $\delta=2\pi(1-D^{1/2})$ where $D$ is where the metric function $A(r)$ plateaus asymptotically.  As one approaches the origin, there is no singularity and the curvature approaches its highest value. We therefore obtain a smooth spacetime with curvature at the origin and near the core of the vortex, that gradually becomes a flat conical spacetime with angular deficit $\delta$ asymptotically. The Ricci scalar at the origin and the angular deficit basically increase tenfold from $\alpha=10$ to $\alpha=1$. For the case of a vortex in fixed Minkowski spacetime (no gravity), we could not obtain the mass using the first method of subtracting two metrics. However, remarkably, we were still able to extract an integral mass formula for it via a limiting procedure. The mass of the vortex with no gravity and the masses of the vortices for all three different $\alpha$ values, were hardly different from each other (maximum of $1\%$ difference).   

We showed that the logarithmic divergence in the energy of the vortex without gauge fields persists in the presence of gravity. However, with gravity, we approached this issue in a new light by looking at the metric instead of the energy integral formula. What we find is that asymptotically, the metric acquires a logarithmic term of the form $G\,m\ln(R)$, which looks like the $2+1$ dimensional Newtonian gravitational potential with mass parameter $m$. It turns out that $m$ is proportional to $n^2v^2$, a product we previously saw in our integral mass formulas. This leaves us with a very important question: can we realize the Newtonian gravitational potential in $2+1$-dimensional General Relativity by supplementing it with a scalar field and potential and choosing a certain profile that leads to finite energy vortices with mass proportional to $n^2v^2$? One possible insight, stemming from the work in \cite{Cadoni}, is that the scalar field can have an asymptotic value of zero even though this is not the minimum of the potential $V(|\phi|)$ (i.e. the extremum that preserves the $U(1)$ symmetry in contrast to the symmetry breaking minimum). What matters is that it satisfies the weaker BF bound $m^2\,L^2\ge -(d-1)^2/4$. Here $m^2=V''(0)$ where the derivatives are with respect to $|\phi|$, $L$ is the AdS length and $d$ is the number of spacetime dimensions. In our work, $d=3$, $L^2=-1/\Lambda$ and $V(f)= \frac{\lambda}{4}(f^2-v^2)^2$ so that $V''(0)=-\lambda v^2$. The BF bound then reads $\frac{\lambda v^2}{\Lambda} \ge -1$. This is already satisfied with some of the parameters we have used in this work (e.g. $\lambda=1$, $v=1$ and $\Lambda=-1$ saturates the bound, while $\lambda=1$, $v=1$ and $\Lambda=-2$ clearly satisfies the bound). This implies that it is worth investigating what happens (in the absence of gauge fields) when the profile of $f(r)$ starts initially at the VEV $v$ and then asymptotically approaches zero. A quick preliminary analysis shows that near $r=0$ there would be a logarithmic dependence with coefficient proportional to $n^2v^2$ (leading probably to a singularity in the spacetime) but not necessarily a long range logarithm so that the energy might in the end turn out to be finite. This and related scenarios look therefore promising and worth studying in more depth to determine if they are viable options.

\pagebreak
\begin{appendices}
\numberwithin{equation}{section}
\setcounter{equation}{0}

\section{Integral mass representation}
\numberwithin{equation}{section}
\setcounter{equation}{0}

In this section we derive an integral expression over purely matter fields for the ADM mass given by \reff{M2} and \reff{MFlat} for an AdS$3$ and flat background respectively.  The equations of motion \reff{EOMf2} and \reff{EOMa2} both contain the functions $A(r)$ and $A'(r)$ plus matter fields. Substituting $A'(r)$ from one equation into the other, we can solve for $A(r)$ in terms of the matter fields. This yields 
\beq
A(r)=\dfrac{n^2 f a'-r^2 v^2 \lambda  f a'-2 n a f a'+a^2 f a'+r^2 \lambda  f^3 a'+e^2 n r^2 f^2 f'-e^2 r^2 a f^2 f'}{r \left(2 a' f'-r f' a''+r a' f''\right)}\,. 
\eeq{A}
Substituting $A(r)$ above into \reff{EOMB2} and solving for $A'(r)$ yields 
\pagebreak
\begin{align}
\label{APrime}
A'(r)&=-2 \Lambda r +\frac{1}{4 e^2 r \alpha }\Bigg[-e^2 r^2 v^4 \lambda -2 e^2 \left(n^2-r^2 v^2 \lambda -2 n a+a^2\right) f^2-e^2 r^2 \lambda  f^4\\\nonumber&\qquad\qquad-\dfrac{2} {r \left(2 a' f'-r f' a''+r a' f''\right)}\Big(n^2 f a'-r^2 v^2 \lambda  f a'-2 n a f a'+a^2 f a'\\\nonumber&\qquad\qquad\qquad+r^2 \lambda  f^3 a'+e^2 n r^2 f^2 f'-e^2 r^2 a f^2 f'\Big) \left(a'^2+e^2 r^2 f'^2\right)\Bigg]
\end{align}
Integrating the above from $0$ to the computational boundary $R$ yields
\begin{align}
A(R)=-\Lambda R^2 + C -\chi= A_0(R) -\chi
\label{AA}
\end{align} 
where $C=A(0)=A_0(0)$ (which we set to unity in this work) and  
\begin{align} 
\label{Chi}
\chi&= \frac{1}{4 e^2 \alpha }\int_0^R \frac{1}{r}\Bigg[e^2 r^2 v^4 \lambda +2 e^2 \left(n^2-r^2 v^2 \lambda -2 n a+a^2\right) f^2+e^2 r^2 \lambda  f^4\\\nonumber&\quad\quad\quad+\dfrac{2}{r \left(2 a' f'-r f' a''+r a' f''\right)}\Big(n^2 f a'-r^2 v^2 \lambda  f a'-2 n a f a'+a^2 f a'\\\nonumber&\qquad\qquad\qquad+r^2 \lambda  f^3 a' +e^2 n r^2 f^2 f'-e^2 r^2 a f^2 f'\Big) \left(a'^2+e^2 r^2 f'^2\right)\Bigg]\, dr\,.
\end{align}
The ADM mass in an AdS$_3$ background, given by \reff{M2} is 
\beq
M_{AdS_3}= 2 \pi \alpha [A_0(R)-A(R)]= 2 \pi \alpha \chi =I
\eeq{M3}
where $I$ is given by 
\begin{align} 
\label{Int2}
I&= \frac{\pi}{2 e^2}\int_0^R \frac{1}{r}\Bigg[e^2 r^2 v^4 \lambda +2 e^2 \left(n^2-r^2 v^2 \lambda -2 n a+a^2\right) f^2+e^2 r^2 \lambda  f^4\\\nonumber&\quad\quad\quad\dfrac{2}{r \left(2 a' f'-r f' a''+r a' f''\right)}\Big(n^2 f a'-r^2 v^2 \lambda  f a'-2 n a f a'+a^2 f a'\\\nonumber&\qquad\qquad\qquad+r^2 \lambda  f^3 a' +e^2 n r^2 f^2 f'-e^2 r^2 a f^2 f'\Big) \left(a'^2+e^2 r^2 f'^2\right)\Bigg]\, dr\,.
\end{align}
Note that $\alpha$ has cancelled out in $I$. Therefore $M_{AdS_3}=I$ is expressed as an integral over matter profiles with no reference to metrics, Newton's constant or the cosmological constant.

For a vortex under gravity in asymptotically flat spacetime ($\Lambda=0$) we have from \reff{AR} that $A(R)=D$ and from \reff{A0} that $A_0(R)=C=1$. From \reff{AA} we obtain that $D=1-\chi$. Substituting this into \reff{MFlat} we obtain the integral mass representation 
\beq
M_{flat}= 4 \pi\alpha \Big(1- \sqrt{1-\chi}\Big)
\eeq{MFlat2}
where the integral $\chi$ is over matter profiles and given by \reff{Chi}. Note that $M_{flat}$ depends on $\alpha$ in contrast to $M_{AdS_3}$. To obtain the mass for the case of a vortex with no gravity (i.e. fixed Minkowski spacetime) we take the limit of $M_{flat}$ as $\alpha \to \infty$. This yields 
\beq
M_{no-gravity}=\lim_{\alpha\to \infty} 4 \pi\alpha [1- (1-\chi/2 +...)]=2 \pi \alpha \chi=I
\eeq{NoGravity}
where in the above binomial expansion higher order terms denoted by the ellipsis make zero contribution in the limit $\alpha \to \infty$. Therefore $M_{no-gravity}$ is evaluated using the same integral $I$ as $M_{AdS_3}$.     

We end this appendix by rewriting the integral $I$ given by \reff{Int2} in a convenient form that reveals its dependence on $v^2$ and $n^2$. $f(r)$ and $a(r)$ reach asymptotically (plateau at) $v$ and $n$ respectively. We therefore can write $f(r)=v\,f_1(r)$ and $a(r)=n\, a_1(r)$ where $f_1(r)$ and $a_1(r)$ both reach asymptotic values of unity. Define $u=\tfrac{e\,v}{n}\,r$. Then $a'(r)= e\,v a_1'(u)$, $a''(r)=
\tfrac{e^2\,v^2}{n} a_1''(u)$, $f'(r)=\tfrac{e\,v^2}{n}\,f_1'(u)$, $f''(r)=\tfrac{e^2\,v^3}{n^2} \,f_1''(u)$ and $dr=n\,du/(ev)$. Derivatives on functions with subscript `1' are with respect to $u$. Substituting this into the integral $I$ yields
\beq
 I= n^2 v^2 F\big(\frac{\lambda}{e^2}\big)
\eeq{I2}
where the integral $F$ is given by
\begin{align}
&F\big(\frac{\lambda}{e^2}\big)\\
&=\dfrac{\pi}{2}\int_0^{R_1} \frac{1}{u}\Bigg[u^2 \frac{\lambda}{e^2} + f_1^4 u^2\frac{\lambda}{e^2}  + 2 f_1^2 \big((-1 + a_1)^2 - u^2 \frac{\lambda}{e^2} \big) \nonumber\\& \quad+ \frac{2 f_1 \big((a_1^{\prime})^2 + (f_1^{\prime})^2 u^2\big) \Big((1 - a_1) f_1 f_1^{\prime} u^2 + a_1^{\prime} \big((-1 + a_1)^2  + (-1 + f_1^2) u^2 \frac{\lambda}{e^2}\big)\Big)}{u (2 a_1^{\prime} f_1^{\prime} - a_1^{\prime\prime} f_1^{\prime} u + a_1^{\prime} f_1^{\prime\prime} u)}\Bigg] du
\label{F2}
\end{align}  
where $R_1=\frac{e\,v}{n} R$. Here $F$ is a function of $\lambda/e^2$ and a functional of the matter profiles $f_1(u)$ and $a_1(u)$.

\section{Equivalence of two integrals for the no gravity case}
\numberwithin{equation}{section}
\setcounter{equation}{0}

In textbooks that discuss the vortex in fixed Minkowski spacetime (e.g. \cite{Weinberg}), the mass is given by

\beq
I_{no-gravity}= \pi \int_0^R \Big[\dfrac{(a')^2}{e^2\,r} +r\,(f')^2 +\dfrac{(n-a)^2\,f^2}{r}  
+\dfrac{\lambda r}{2}(f^2-v^2)^2\Big]dr
\eeq{INoGrav}
This looks different from $I$ given by \reff{Int2}. We now show using the equations of motion for the no gravity case i.e.
\begin{align}
&r f \left(-\frac{(n-a)^2}{r^2}+\lambda  \left(v^2-f^2\right)\right)+f'+r f''=0\\
&e^2 r (n-a) f^2-a'+r a''=0
\end{align}
that the two are equivalent. Solving for $f''$ and $a''$ we obtain
\begin{align}
&f''=\frac{n^2 f-r^2 v^2 \lambda  f-2 n a f+a^2 f+r^2 \lambda  f^3-r f'}{r^2}\\
&a''=\frac{-e^2 n r f^2+e^2 r a f^2+a'}{r}\,.
\end{align}
Substituting $f''$ and $a''$ above into the quantity
\beq
\dfrac{(n^2 f a'-r^2 v^2 \lambda  f a'-2 n a f a'+a^2 f a'+r^2 \lambda  f^3 a' +e^2 n r^2 f^2 f'-e^2 r^2 a f^2 f')}{r \left(2 a' f'-r f' a''+r a' f''\right)} 
\eeq{Focus}
that appears in the integral \reff{Int2} yields unity. The integral $I$ then reduces to
\begin{align} 
&\frac{\pi}{2 e^2}\int_0^R \frac{1}{r}\Bigg[e^2 r^2 v^4 \lambda +2 e^2 \left(n^2-r^2 v^2 \lambda -2 n a+a^2\right) f^2+e^2 r^2 \lambda  f^4 +2\left(a'^2+e^2 r^2 f'^2\right)\Bigg]\, dr\nonumber\\
&=\pi \int_0^R \Big[\dfrac{(a')^2}{e^2\,r} +r\,(f')^2 +\dfrac{(n-a)^2\,f^2}{r}  
+\dfrac{\lambda r}{2}(f^2-v^2)^2\Big]dr\nonumber\\
&=I_{no-gravity}\,.
\end{align}
We therefore see that the two integrals are equivalent in the case of no gravity.
\end{appendices}

%---------------------------------------------------------------------------------------
\section*{Acknowledgments}
A.E. acknowledges support from a discovery grant of the National Science and Engineering Research Council of Canada (NSERC). I thank the referees for their valuable comments. I thank G-d for giving me the strength and wisdom to complete this project.

\end{document}